\documentclass[11pt,a4paper]{article}

\usepackage[utf8]{inputenc}
\usepackage[T1]{fontenc}
\usepackage{amsmath,amssymb}
\usepackage{bm}
\usepackage{graphicx}
\usepackage{booktabs}
\usepackage{hyperref}
\usepackage{url}
\usepackage[numbers,sort&compress]{natbib}
\usepackage{authblk}
\usepackage[margin=2.5cm]{geometry}
\usepackage{caption}
\usepackage{float}
\usepackage{enumitem}
\usepackage{xcolor}

\hypersetup{colorlinks=true, linkcolor=blue, citecolor=blue, urlcolor=blue}
\newcommand{\Us}{U^{*}}
\newcommand{\UsF}{U^{*}_{F}}
\newcommand{\MR}{M_{R}}
\newcommand{\vect}[1]{\bm{#1}}
\newcommand{\doi}[1]{\href{https://doi.org/#1}{doi:#1}}
\newcommand{\orcid}[1]{\textsuperscript{\href{https://orcid.org/#1}{\scriptsize ORCID}}}

\title{\textbf{Compressible Aerodynamics and Rigid-Body Support Motion in
Post-Flutter Piezoelectric Energy Harvesting from a Pitch-Plunge-Flap
Aerofoil}}

\author[1]{Nikolaos~D.~Tantaroudas\orcid{0000-0001-6727-1014}\thanks{Corresponding
author: \texttt{nikolaos.tantaroudas@iccs.gr}.}}
\author[2]{Ilias~Karachalios\orcid{0009-0000-8581-8335}}
\author[3]{Andrew~J.~McCracken}

\affil[1]{Institute of Communications and Computer Systems (ICCS), National
Technical University of Athens, Iroon Polytechniou 9, 15773 Zografou, Greece}
\affil[2]{Department of Environmental Sciences, University of Thessaly, Gaiopolis
Campus, Larissa-Trikala Ring Road, 41500 Larissa, Greece}
\affil[3]{DASKALOS APPS, 183 Rue de l'Abb\'{e} Griffon, 01960 P\'{e}ronnas,
France}

\date{}

\begin{document}
\maketitle

\begin{abstract}
A piezoelectric transducer on a pitch-plunge aerofoil with a finite-mass trailing-edge flap
turns the limit cycle that follows flutter into electrical power, and two modelling changes to
that harvester have been characterised separately. Replacing incompressible strip theory with
the compressible Euler equations was found to reduce the stabilising effect of a pitch-mounted
transducer on the flutter boundary, where the electrical time constant meets the flutter
frequency, and to raise the harvested power. Letting the section's suspension react against a
fuselage free to translate, rather than against rigid ground, was found to do the same two
things by about the same amounts. Neither study could say what the two do together, because
neither had both. Here they are put in one model, which turns out to need no new mesh motion
in the flow solver at all, the wall displacement conjugate to the support's translation being
identically the one conjugate to the plunge, so that the whole extension is a motion map and a
force map at the aerodynamic interface. The two changes are found to compose on one
quantity everywhere and on the other only in one of two regimes. On the harvested power they are multiplicative, so that the two
published corrections may simply be multiplied. On the flutter boundary they more than
compound, and at a strong enough electromechanical coupling the combination reverses the sign
of the transducer's effect at the load resistance at which both companion papers do their
harvesting, where neither change alone reverses it. A reversed sign is not itself new, the
compressible companion paper having found one at a tenfold smaller load resistance, but the
free support is shown to widen the band of resistance over which the two aerodynamic models
disagree in sign, carrying its upper edge out by about a factor of three. Multiplying the two
published corrections there gives the wrong sign. Sweeping the resistance on both mountings
locates where that failure begins and where it stops. Above the resistance at which the
electrical pole meets the flutter frequency, where the transducer acts as an added stiffness,
the product of the two published corrections predicts the measured shift to within seven parts
in a thousand on either mounting. Below it, where the transducer acts as a phase-dependent
damper and where both companion papers reported their effects, the product is wrong by a factor
of between one and a half and fourteen, and on the pitch mounting at the harvesting resistance
by a sign. The same transducer
is further shown to lower the boundary while flattening the growth of the instability above
it, so that it is destabilising in the sense that decides where the cycle begins and
stabilising in the sense that decides how fast it grows. Whether studies that characterise one
modelling change at a time may be composed by multiplying their corrections therefore depends
on the quantity, and for the flutter boundary on the regime the transducer is working in,
rather than on the changes themselves.
\end{abstract}


\section{Introduction}
\label{sec:intro}

Flutter is ordinarily a boundary to design away from, and the limit cycles beyond it are a
fatigue problem. The same instability is also a steady conversion of flow energy into
structural motion, and a piezoelectric element embedded in the structure turns part of that
motion into electricity. Erturk and co-workers measured $10.7$~mW at the linear flutter speed
of a piezoaeroelastic airfoil and observed that the transduction itself moves the flutter
boundary~\citep{erturk2010}, which is the effect this series follows. A linear system at its
boundary has no bounded amplitude, so devices work above it on a cycle fixed by a
nonlinearity. Power has been taken from the limit cycles of a beam in axial
flow~\citep{dunnmon2011}, raised by combining free play with a hardening
spring~\citep{sousa2011}, compared between piezoelectric and electromagnetic transduction on
airfoil devices~\citep{demarqui2013}, and measured on a flutter-driven harvester built for the
purpose~\citep{bryant2011}. The field is reviewed in~\citep{abdelkefi2016}.

Three companion papers precede this one, and together they vary two things about the same
harvester. A piezoelectric transducer is embedded in a two-dimensional pitch-plunge aerofoil
with a finite-mass trailing-edge flap, and the limit cycle that follows flutter is converted
into electrical power. The first~\citep{tantaroudas2026harvest} used incompressible strip
theory and mounted the section on rigid ground, and found that the degree of freedom carrying
the transducer sets both the sign of the shift of the flutter boundary and the harvested
power. The second~\citep{tantaroudas2026cfd} replaced the aerodynamics with the compressible
Euler equations and kept the rigid mounting. The third~\citep{tantaroudas2026rigid} kept strip
theory and took away the ground, letting the section's suspension react against a fuselage
that is itself free to translate, so that the coupled system is free-free.

Each of the last two found the same thing in the same place, and neither could say what the
two do together. Where the electrical time constant $1/(\Us r)$ meets the flutter frequency,
the pitch mounting's stabilising shift of the flutter boundary is roughly halved by a
compressible flow, and it is roughly halved again by a free support. The two effects have the
same sign and comparable size. Whether they compound, so that a compressible flow on a free
support leaves a quarter of the shift, or whether they are two descriptions of one mechanism
and so partly the same effect, is not a question either study could answer, because neither had both effects in
one model. The present paper has both.

The free-free coupling between a wing's elastic degrees of freedom and a body's rigid-body ones
is the coupling that makes aeroelastic-flight-dynamic models of flexible aircraft
distinctive~\citep{tantaroudas2026coupled}, and reduced-order models of that line have been used
to identify worst-case gusts rapidly~\citep{tantaroudas2026gust}.

It also settles a practical question. Coupling a flow solver to the free-free section might be
expected to require a four-degree-of-freedom mesh motion. It does not, for a reason given in
Section~\ref{sec:model} that makes the extension almost free, and the point is worth stating on
its own because it removes what looks like the main obstacle to putting high-fidelity
aerodynamics on a free-free aeroelastic harvester.

\section{Model}
\label{sec:model}

\subsection{The section, the support and the transducer}

The structure is that of the third companion paper and is not restated in full. The
generalised coordinates are $\vect{q} = (\xi,\alpha,\delta,y)^{\mathsf T}$, the wing's plunge
relative to the fuselage, its pitch, the flap deflection and the fuselage's own translation.
The wing's absolute plunge is $\xi + y$. The mounting enters through one dimensionless group,
the ratio $\MR = m/M_f$ of the wing section's mass to the fuselage's, and $\MR = 0$ is a
fuselage of infinite mass, which is the rigidly supported section of the first two papers.
The transducer is a lumped element across one structural degree of freedom with coupling
$\chi$ and dimensionless load resistance $r$, and the three mountings are the plunge, the
pitch and the flap.

\subsection{Why a flow solver needs no fourth mesh motion}
\label{sec:mesh}

The aerodynamic model, whichever it is, is a model of a two-dimensional section loaded by its
\emph{absolute} plunge, its pitch and its flap deflection. It has three coordinates and not
four. The fourth generalised coordinate reaches it through the motion map
\begin{equation}
\vect{T} =
\begin{bmatrix} 1 & 0 & 0 & 1 \\ 0 & 1 & 0 & 0 \\ 0 & 0 & 1 & 0 \end{bmatrix},
\qquad
\vect{T}\vect{q} = (\xi + y,\ \alpha,\ \delta)^{\mathsf T},
\label{eq:Tmap}
\end{equation}
and the generalised force comes back through $\vect{R}$, which is $\vect{T}^{\mathsf T}$ with
its fuselage row scaled by $\MR$, that being the fuselage equation's own normalisation. Every
aerodynamic operator is $\vect{R}(\cdot)\vect{T}$.

For strip theory that is a statement about matrices. For a flow solver it is a statement about
the mesh, and it is what makes the extension cheap. The loads are computed as the virtual work
of the wall forces through the wall displacement conjugate to each coordinate,
$Q_j = \sum_{\mathrm{faces}} \vect{F}\cdot\partial \vect{x}/\partial q_j$, and for the
fuselage translation that displacement is
\begin{equation}
\frac{\partial \vect{x}}{\partial y} = \left(0,\ -\tfrac{1}{2}\right)
= \frac{\partial \vect{x}}{\partial \xi},
\label{eq:arms}
\end{equation}
identically and at every state, a rigid vertical translation either way, because moving the
fuselage moves the wing exactly as plunging it does. The flow solver therefore still sees three
coordinates, its mesh motion is unchanged, and the fourth coordinate is carried by
Eq.~\eqref{eq:Tmap} alone.

Two consequences follow, and both are used. The generalised aerodynamic force on the fuselage
is $\MR$ times the force on the plunge, exactly, so nothing new has to be integrated over the
surface. And a rigid translation of the whole system leaves the flow solver's state untouched
in mesh coordinates, the cell volumes, the face normals and the free stream all being
unchanged, so the aerodynamic block of the free-free system's null vector is \emph{zero} where
strip theory's lag states, being deviations driven by the displacement, are not.

\subsection{The compressible aerodynamics}

The flow model is the second companion paper's and is likewise not restated. The conservative
variables of every cell of a NACA O-mesh are marched in the same reduced time as the
structure, pitch and plunge by rigid mesh motion and the flap by radial-basis-function
deformation, with the wall forces taken from the flux the residual actually applies through
the wall faces rather than from a reconstructed wall pressure. There is no apparent-mass term,
a flow solver's load being a surface integral over the flow state and never a function of
$\vect{q}''$, so the assembler inverts the structural mass alone. The Mach number is a deck
input and the aerodynamics in reduced time does not depend on the reduced velocity, exactly as
for strip theory.

Collecting the structure, the flow and the transducer gives a coupled autonomous system whose
aerodynamic block is the flow field rather than eight lag variables.

\section{Implementation and validation}
\label{sec:impl}

\subsection{What was carried over and what was written}

The solver is the third companion paper's, whose aerodynamics was the unsteady strip theory of
the pitch-plunge-flap section of Lee and co-workers~\citep{lee1999} as implemented
in~\citep{tantaroudas2015phd}, with the second paper's flow provider attached behind the same
seam. The two maps of
Section~\ref{sec:mesh} are stated once and shared. Strip theory applies them to its operators
when it builds them, the flow solver applies them at its interface, and below that interface
every line is the three-degree-of-freedom solver's, unchanged.

Two things did have to be written. The rigid-body null vector is now supplied by the
aerodynamic model rather than solved out of its linearisation, since for a flow solver that
linearisation costs a whole Jacobian assembly and the answer is zero. And the flutter search
is bisection when a spectrum is cheap and the Illinois rule on the same bracket when it is
not, one spectrum of the coupled flow model being the better part of a minute against
microseconds for strip theory. On strip theory the two searches agree to $10^{-7}$, and every
strip-theory boundary of the third paper is reproduced bit for bit.

\subsection{Validation}

The compressible model is validated the way the strip-theory one was, by reproducing a
published companion result in the heavy-fuselage limit, and it is done twice on two different
things.

\textbf{The baseline.} With no transducer and $\MR = 0$, the flutter boundary at $M = 0.10$ on
the ninety-six by thirty-two mesh is $\UsF = 6.288873$ against its $6.288873$, a difference of
$2.9\times10^{-7}$, at a frequency of $0.08352$ against that paper's $0.0835$. That exercises
the aerodynamics, the mesh motion and the coupling to the structure.

\textbf{The transducer.} Not one point but the companion paper's boundaries at
$\chi = 0.50$ on both mountings, reproduced on a rigid support and compared against that
paper's own recorded values rather than against its printed table, which is rounded to four
figures and would hide agreement at the eleventh
(Table~\ref{tab:cfd_reproduction}). Every one of them agrees to
$4\times10^{-11}$, which is the tolerance of the two searches rather than a property of the
physics, and that holds across all $8$ of its recorded boundaries, four on each mounting and
spanning three decades of load resistance. That exercises the harvester, which the baseline does not, and it is
the more telling of the two. The electrical row and column, the mounted degree of freedom and
the reference power all have to be right for it, and a four-degree-of-freedom layout puts
every one of them in a different place.

\begin{table}[H]
\centering
\caption{The companion paper's boundaries at $\chi = 0.50$ on a rigid support, against the
four-degree-of-freedom solver with its support frozen.}
\label{tab:cfd_reproduction}
\small
\begin{tabular}{llrrr}
\toprule
mounting & $r$ & companion paper & this work & difference \\
\midrule
none & baseline & 6.288873 & 6.288873 & 2.9e-07 \\
pitch & 0.1 & 6.281884 & 6.281884 & 3.9e-11 \\
 & 1 & 6.411568 & 6.411568 & 3.0e-11 \\
 & 10 & 7.899890 & 7.899890 & 3.2e-11 \\
 & 100 & 7.949124 & 7.949124 & 3.2e-11 \\
\addlinespace
plunge & 0.1 & 6.293348 & 6.293348 & 3.6e-11 \\
 & 1 & 6.261701 & 6.261701 & 2.0e-11 \\
 & 10 & 6.084875 & 6.084875 & 2.0e-11 \\
 & 100 & 6.056708 & 6.056708 & 2.0e-11 \\
\bottomrule
\end{tabular}

\end{table}

Both are reproductions rather than agreements, since the $\MR \to 0$ limit is exact by
construction. What they test is that the construction was carried out correctly.

\textbf{The transducer at zero coupling.} With the harvester enabled but $\chi = 0$ the
leading eigenvalue is unchanged to $2.3\times10^{-16}$. A transducer that exerts no force must
leave the mechanical spectrum exactly where it was, and on a four-degree-of-freedom layout
that is a statement about where the electrical row and column were assembled.

\textbf{Reproducibility.} One configuration was computed twice over, in two separate processes
started at different times and running the linear algebra on different numbers of threads. The two boundaries agree to $1.9\times10^{-12}$, so the search is not sensitive to the order in which the reductions happen, which for an iterative eigenvalue solve on a sparse
operator of twelve thousand states is worth knowing rather than assuming.

\textbf{The time step.} The strip-theory study marches at a step of $0.05$ and the flow model
at $0.2$, the latter being what an implicit march of a stiff provider can afford, so a
difference in harvested power between the two models could in principle be partly a difference
of step. Marching the strip-theory cycle at both steps changes its mean power by
$1.5\times10^{-4}$ of itself, three orders below the differences reported here, so the
comparison is a comparison of aerodynamics.

\textbf{The mesh.} The two results above are on the ninety-six by thirty-two mesh of the
companion paper. At the anchor support mass the boundary is $6.261609$ on sixty-four by
twenty-four, $6.266810$ on the paper's mesh and $6.271649$ on one hundred and twenty-eight by
forty-eight, a spread of $0.160\%$ across a fourfold refinement with the paper's mesh within
$0.077\%$ of the finest. The support mass does not change what the companion paper found about
the mesh, which is what this check is for, since every difference the paper reports is larger
than that by an order of magnitude.

\begin{table}[H]
\centering
\caption{The flutter boundary against the support mass, strip theory beside the compressible
flow.}
\label{tab:cfd_boundary}
\small
\begin{tabular}{rrrrrr}
\toprule
$M_R$ & \multicolumn{2}{c}{strip theory} & \multicolumn{2}{c}{Euler, $M = 0.10$} & difference \\
\cmidrule(lr){2-3}\cmidrule(lr){4-5}
 & $\Us_F$ & shift [\%] & $\Us_F$ & shift [\%] & in $\Us_F$ [\%] \\
\midrule
0.00 & 6.3246 & +0.000 & 6.2889 & +0.000 & -0.565 \\
0.10 & 6.3179 & -0.105 & 6.2817 & -0.114 & -0.574 \\
0.20 & 6.3111 & -0.213 & 6.2743 & -0.231 & -0.583 \\
0.30 & 6.3042 & -0.323 & 6.2668 & -0.351 & -0.593 \\
0.50 & 6.2898 & -0.550 & 6.2512 & -0.599 & -0.614 \\
0.70 & 6.2747 & -0.789 & 6.2348 & -0.860 & -0.636 \\
1.00 & 6.2507 & -1.168 & 6.2086 & -1.276 & -0.673 \\
\bottomrule
\end{tabular}

\end{table}

\section{Results}
\label{sec:results}

The pitch mounting at $\chi = 0.50$ and $r = 1$ is where both effects are known to act. Strip
theory on a rigid support gives $+4.06\%$, and each of the two changes roughly halves it. With
a compressible flow on a rigid support the shift is $+1.951\%$, which is the second companion
paper's number, and with strip theory on a free support $+1.94\%$, which is the third's. The cell
neither could compute is the one with both, and both together leave $-0.533\%$.

The sign is reversed at this load resistance. Both past studies report a positive shift
there, and so does strip theory on a free support, and the compressible flow on a free support
does not. That is at $\chi = 0.50$, and Section~\ref{sec:coupling} shows the reversal needs a
coupling of that order.

It should be said at once that a reversed sign is not by itself new. The past paper using compressible flow found one, at $r = 0.1$ on a rigid support, and reported it as a small negative shift of $-0.11\%$ where strip theory gives $+0.385\%$. This work reproduces that
number as $-0.111\%$ and finds the same sign at $r = 0.3$, at $-0.232\%$. What is new is
Section~\ref{sec:band}, which shows that the free support does not create the reversal but
\emph{widens} it, carrying its upper edge from between $r = 0.3$ and $r = 1$ up to between
$r = 1$ and $r = 3$, so that it now covers the anchor load resistance at which both companion
papers do their harvesting.

Simple additivity of the two shifts would predict $-0.17\%$ at this resistance, so the two
rather more than compound, and Section~\ref{sec:coupling} shows that they do so at both
couplings tested. The important point is not the arithmetic. It is that a designer who took
either single-effect study and applied it to a section that has both would get the sign of the
transducer's effect on stability wrong at the load resistance the harvesting is done at, and
not merely its size.

\begin{table}[H]
\centering
\caption{The pitch mounting's shift of the flutter boundary at the coupling and load
resistance where both effects act. The bottom right cell is the one no previous study had.}
\label{tab:open}
\small
\begin{tabular}{lrr}
\toprule
aerodynamics & rigid support & free support, $M_R = 1$ \\
\midrule
incompressible strip theory & +4.06 & +1.94 \\
compressible Euler, $M = 0.10$ & +1.95 & -0.53 \\
\bottomrule
\end{tabular}

\end{table}

\begin{figure}[H]
\centering
\includegraphics[width=0.62\textwidth]{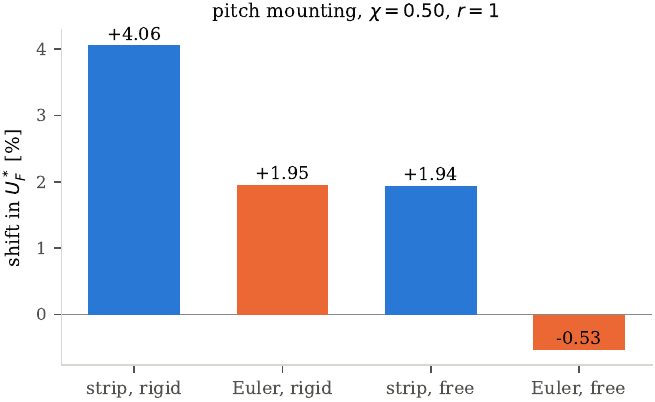}
\caption{The same four values. Each effect roughly halves the stabilisation on its own and the
two together carry it past zero.}
\label{fig:open}
\end{figure}

\subsection{The same result without the root-finder}

A bisection reports where a curve crosses zero and says nothing about the curve, so a result
that reverses a sign should not rest on one. The growth rate was therefore taken at two fixed
reduced velocities either side of the baseline boundary, with the transducer off and on, which
compares the two systems directly and has no root-finder in it.

At $\Us = 6.15$, below both boundaries, the leading real part is $-2.664\times10^{-3}$ without
the transducer and $-2.053\times10^{-4}$ with it. The transducer raises the growth rate, which
is the reversal, measured directly. Fitting the two points of each branch puts their zero
crossings at $6.2088$ and $6.1745$, against the $6.208648$ and $6.175578$ the bisection found,
so the two methods agree.

The same two points show something the boundary alone hides. The transducer's branch crosses
zero earlier and also rises about five times more slowly, its slope being $0.0084$ against
$0.0453$ per unit reduced velocity. So the same transducer that lowers the boundary reduces
the growth rate above it, and at $\Us = 6.25$, above both boundaries, the system with the
transducer is the more stable of the two. Both statements are true of the same device and a
harvester paper should make both, since the first governs where the cycle begins and the
second how fast it grows.

\subsection{Load Resistance}
\label{sec:band}

A sign reversal at one load resistance would be a curiosity. Sweeping the resistance with the
pitch mounting in the compressible flow, on both supports, shows it is a band, gives it an edge
with a physical meaning, and shows what the free support actually does to it.

On a free support the shift is $-0.435\%$ at $r = 0.1$, deepens to $-1.197\%$ at $r = 0.3$, is
still negative at $-0.533\%$ at $r = 1$, and has turned to $+14.499\%$ by $r = 3$. On a rigid
support the same sweep gives $-0.111\%$, $-0.232\%$, $+1.951\%$ and $+16.053\%$. Strip theory
is positive at every one of those on either support, so the two aerodynamic models disagree in
sign across a band of load resistance rather than at a point.

The free support widens that band. On a rigid support the shift changes sign between $r = 0.3$
and $r = 1$, on a free support between $r = 1$ and $r = 3$, so the upper edge moves out by
about a factor of three and the band comes to include the anchor resistance $r = 1$. The
reversal is deepest near $r = 0.3$ on either support and shallows towards $r = 0.1$, as it
must, since a short-circuited transducer exerts no force and both models return to the
baseline together.

The upper edge is where the transducer changes character. The electrical pole sits at
$-1/(\Us r)$ and the flutter frequency of this system is $0.085362$, so the two are equal at
$r = 1.89$, which lies between the last negative point and the first positive one. Below that
resistance the electrical relaxation is faster than the flutter oscillation and the transducer
acts as a phase-dependent damper, whose effect depends on where the flutter mode sits and on
its frequency. Above it the relaxation is slower, the transducer acts as an added stiffness,
and the shift is large and positive in both models. The reversal belongs entirely to the
damping regime, which is also the regime in which both companion papers reported their
effects.

That is worth stating as the answer to a question those papers could only gesture at. Each
described its effect as appearing where the electrical time constant meets the flutter
frequency. It can now be said quantitatively that the disagreement occupies the band in which
the electrical pole is the faster of the two, and closes when it is not.

\begin{figure}[H]
\centering
\includegraphics[width=\textwidth]{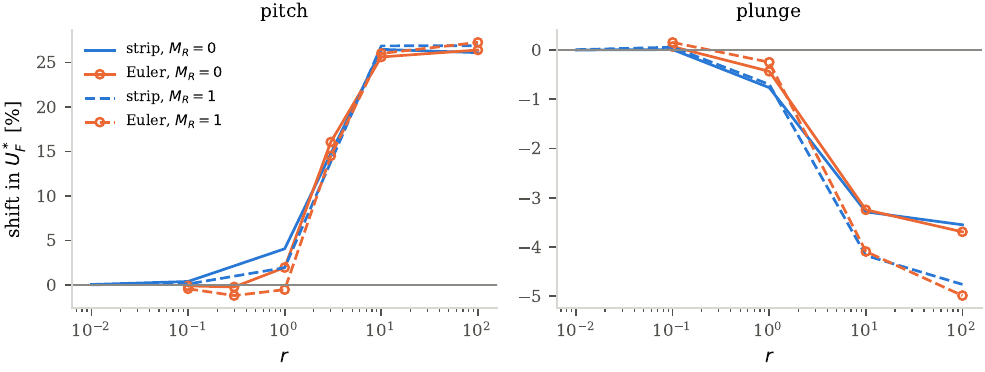}
\caption{The transducer's shift of the boundary against load resistance in both aerodynamic
models, on a rigid and on a free support, one panel per mounting.}
\label{fig:cfd_resistance}
\end{figure}

\subsection{Where the two corrections do multiply}
\label{sec:compose}

The same sweep, read across both mountings, answers a sharper question than whether the sign
reverses. Table~\ref{tab:cfd_compose} puts beside each measured shift the one a reader holding
the two companion papers would predict, by taking the rigidly supported strip-theory value and
multiplying it by the factor each change applies on its own.

The prediction is not uniformly wrong. Above the resistance at which the electrical pole meets
the flutter frequency, the $r = 1.89$ of Section~\ref{sec:band}, it is almost exactly right on
both mountings. At $r = 10$ and $r = 100$ the product gives $+25.993\%$ and $+27.196\%$ on the
pitch mounting against a measured $+26.011\%$ and $+27.251\%$, and $-4.119\%$ and $-4.952\%$ on
the plunge mounting against a measured $-4.091\%$ and $-4.986\%$. The largest of those four
disagreements is seven parts in a thousand.

Below that resistance it fails on both mountings. On the plunge the product is too large by a
factor of $2.6$ at $r = 0.1$ and $1.6$ at $r = 1$, both times with the sign right. On the pitch
it is worse, and wrong the other way, an order of magnitude too small at $r = 0.1$, where it
predicts $-0.030\%$ against a measured $-0.435\%$, and of the wrong sign at $r = 1$, where it
predicts $+0.931\%$ against a measured $-0.533\%$. The plunge mounting never changes sign anywhere in the sweep, both models putting it
slightly positive at $r = 0.1$ and negative at every larger resistance on either support, so the
reversal of Section~\ref{sec:band} belongs to the pitch mounting while the failure to compose
belongs to both.

Composability is therefore not a property of the pair of modelling changes, nor of the mounting,
but of the regime the transducer is working in. Where it acts as an added stiffness the two
changes are independent and their corrections multiply. Where it acts as a phase-dependent
damper they are not, because both changes move the flutter mode and its frequency and the
damping depends on both. That is the same boundary, and the same mechanism, as the sign reversal
of Section~\ref{sec:band}, and it places the failure squarely in the regime both companion
papers chose to report their effects in.

\begin{table}[H]
\centering
\caption{The shift measured with both changes present, against the product of what each change
does alone, at four load resistances on each mounting. The electrical pole meets the flutter
frequency at $r = 1.89$.}
\label{tab:cfd_compose}
\small
\begin{tabular}{lrrrrrr}
\toprule
mounting & $r$ & strip, rigid & Euler, rigid & strip, free & Euler, free & product \\
\midrule
pitch & 0.1 & +0.385 & -0.111 & +0.105 & -0.435 & -0.030 \\
 & 1 & +4.058 & +1.951 & +1.936 & -0.533 & +0.931 \\
 & 10 & +26.470 & +25.617 & +26.860 & +26.011 & +25.993 \\
 & 100 & +26.100 & +26.400 & +26.887 & +27.251 & +27.196 \\
\addlinespace
plunge & 0.1 & +0.010 & +0.071 & +0.059 & +0.156 & +0.411 \\
 & 1 & -0.763 & -0.432 & -0.697 & -0.246 & -0.395 \\
 & 10 & -3.285 & -3.244 & -4.172 & -4.091 & -4.119 \\
 & 100 & -3.548 & -3.692 & -4.759 & -4.986 & -4.952 \\
\bottomrule
\end{tabular}

\end{table}

\subsection{How strong the coupling must be}
\label{sec:coupling}

The reversal is not unconditional, and the condition is the coupling. Repeating the whole
comparison at $\chi = 0.20$, the coupling the limit-cycle studies use, gives $+0.267\%$ rather
than a negative shift. Strip theory on the same free support gives $+1.739\%$, so the two
changes still reduce the stabilisation heavily, but they do not carry it past zero. At
$\chi = 0.50$ they do.

Having both couplings measured in all four configurations says more than that, and it says the
part that does not depend on the coupling. Writing each change as a factor on the rigidly
supported strip-theory shift, the compressible flow multiplies it by $0.477$ at $\chi = 0.20$
and by $0.481$ at $\chi = 0.50$, which is the same factor to three places. The free support
multiplies it by $0.686$ and by $0.477$, so its effect strengthens as the coupling grows. And
at both couplings the two together undershoot the product of their factors, by
$+0.267\%$ against a predicted $+0.829\%$ at the weaker coupling and by $-0.533\%$ against a
predicted $+0.931\%$ at the stronger.

So the super-compounding is the robust finding and the sign reversal is its consequence at a
strong enough coupling. A harvester designer taking the two published corrections and
multiplying them would under-predict the loss of stabilisation at either coupling, and would
additionally get the sign wrong at the stronger one. 
Sweeping the coupling locates the crossing. The shift is $+0.2672\%$ at $\chi = 0.20$,
$+0.0952\%$ at $0.28$, $-0.0938\%$ at $0.35$, $-0.2977\%$ at $0.42$ and $-0.5326\%$ at
$0.50$, falling monotonically throughout, so it changes sign between $0.28$ and $0.35$, at
$\chi = 0.315$ by linear interpolation between the two. A
transducer on this section, in this flow, on this support and at this load resistance
stabilises the boundary below that coupling and destabilises it above.

This also explains why neither past study addressed this. Each reports its largest
effects at $\chi = 0.50$ and each finds a halving there. It is only with both changes present
at that coupling that the halving becomes a reversal.

\subsection{The support mass on its own, in a compressible flow}

\begin{figure}[H]
\centering
\includegraphics[width=0.86\textwidth]{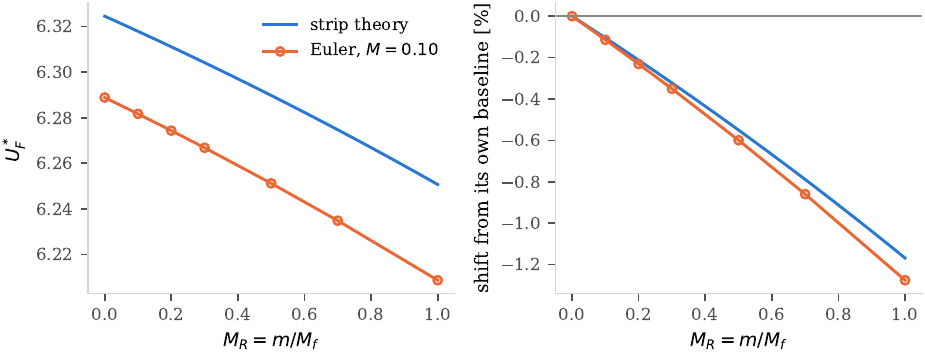}
\caption{The flutter boundary against the support mass in both aerodynamic models, and the
shift of each from its own rigidly supported value.}
\label{fig:cfd_boundary}
\end{figure}

With no transducer the support mass is destabilising in a compressible flow as it is in strip
theory, and slightly more so at every support mass tested (Table~\ref{tab:cfd_boundary},
Figure~\ref{fig:cfd_boundary}). The boundary falls $1.276\%$ over the same range of support mass
across which strip theory gives $1.168\%$. The two effects on the baseline therefore also have
the same sign, and a compressible flow and a free support each move the boundary down.

That is worth separating from the result above. On the \emph{baseline} the two changes are
small and merely additive. It is on the transducer's \emph{shift} that they compound enough to
reverse a sign, and the reason is that the shift at $r = 1$ is a phase-dependent damping rather
than an added stiffness, so it is sensitive to the flutter mode's shape and frequency in a way
the baseline is not. Both changes alter that mode, and the transducer's own effect on it is
large here. Switching the transducer on at $\Us = 6.15$ moves the flutter frequency from
$0.0846$ to $0.0680$.

\section{The limit cycle and the harvested power}
\label{sec:power}

The flutter boundary is where the cycle begins. What a harvester delivers is set by the cycle
itself, and the two effects behave differently there.

At the anchor velocity with the plunge-mounted transducer at $\chi = 0.20$ and $r = 1$, the
mean power on the common mechanical reference is $1.799\times10^{-5}$ for strip theory on a
rigid support. Making the flow compressible raises it to $2.431\times10^{-5}$, a gain of
$35.1\%$, which is the second companion paper's finding that strip theory under-predicts the
power of a compressible section. Freeing the support instead raises it to
$2.434\times10^{-5}$, a gain of $35.3\%$, which is the third's. Doing both gives
$3.283\times10^{-5}$.

That is a gain of $82.5\%$ over the rigidly supported strip-theory section, against the
$82.8\%$ the two gains would give if they simply multiplied. On the harvested power the two
effects are therefore multiplicative to within three tenths of a percentage point in $82.5$,
which is four parts in a thousand. Each change raises the amplitude of the cycle by about a
third and neither spends any of what the other would have gained. The residual is not claimed
to be zero, only to be small beside the effects themselves and beside the disagreement on the
flutter boundary, which is a change of sign.

That is the paper's second result, and it is a contrast rather than a similarity. On the
harvested power the two changes compose in the simplest way there is, so a designer may
multiply the two published corrections and be right. On the transducer's shift of the flutter
boundary at this resistance they do not compose, and at a coupling strong enough to reverse the
sign, multiplying the two published corrections is wrong by that sign. The same pair of
modelling changes is therefore benign for one quantity and treacherous for the other at the one
operating point, and which it is cannot be guessed from the changes themselves.
Section~\ref{sec:compose} shows that the treachery is confined to one regime of load resistance
rather than general.

The reading is that power is set by the amplitude of the cycle, which both changes raise by
making the path between the air and the transducer more compliant, and amplitude gains of a
third compound as gains do. The shift of the boundary at $r = 1$ is not an amplitude. It is a
phase-dependent damping that depends on where the flutter mode sits and on its frequency, and
both changes move both, so their effects there are not independent.

\begin{table}[H]
\centering
\caption{The limit cycle at the anchor velocity, strip theory beside the compressible flow,
on a rigid and on a free support.}
\label{tab:cfd_lco}
\small
\begin{tabular}{rlrrrrrr}
\toprule
$M_R$ & aerodynamics & $\alpha$ & $\xi$ & $y$ & $v$ & $\langle\bar p\rangle$ & period \\
\midrule
\multicolumn{8}{l}{\itshape $\chi = 0.20$, $r = 1$, plunge} \\
0.00 & strip & 0.0524 & 0.1318 & 0.0000 & 0.0670 & 1.799e-05 & 74.32 \\
 & Euler & 0.0534 & 0.1546 & 0.0000 & 0.0778 & 2.431e-05 & 75.23 \\
0.30 & strip & 0.0530 & 0.1373 & 0.0049 & 0.0700 & 1.967e-05 & 73.95 \\
 & Euler & 0.0541 & 0.1610 & 0.0059 & 0.0814 & 2.658e-05 & 74.85 \\
1.00 & strip & 0.0547 & 0.1513 & 0.0176 & 0.0779 & 2.434e-05 & 73.05 \\
 & Euler & 0.0557 & 0.1772 & 0.0211 & 0.0905 & 3.283e-05 & 73.87 \\
\multicolumn{8}{l}{\itshape no transducer} \\
0.00 & strip & 0.0516 & 0.1372 & 0.0000 & 0.0000 & 0.000e+00 & 76.15 \\
 & Euler & 0.0531 & 0.1639 & 0.0000 & 0.0000 & 0.000e+00 & 77.31 \\
0.30 & strip & 0.0522 & 0.1436 & 0.0051 & 0.0000 & 0.000e+00 & 75.89 \\
 & Euler & 0.0538 & 0.1717 & 0.0063 & 0.0000 & 0.000e+00 & 77.03 \\
1.00 & strip & 0.0539 & 0.1601 & 0.0187 & 0.0000 & 0.000e+00 & 75.23 \\
 & Euler & 0.0556 & 0.1922 & 0.0232 & 0.0000 & 0.000e+00 & 76.37 \\
\bottomrule
\end{tabular}

\end{table}

\textbf{Where the compressible flow puts the extra motion.} It is almost all in the plunge.
Over every pair of cycles measured, with the transducer and without and on both supports, the
compressible cycle's plunge is between $17.1\%$ and $20.1\%$ larger than strip theory's while
its pitch is between $1.9\%$ and $3.1\%$ larger. The two models therefore disagree much less
about how far the section twists than about how far it heaves, which is consistent with the
transducer's own preference, the plunge mounting outperforming the pitch mounting by a factor
of about four in both models, and it being the plunge that the compressible flow enlarges.

The support motion itself is the quantity the two models differ on most in relative terms. On
a free support the compressible flow drives the fuselage between a fifth and a quarter harder,
over all four pairs that have a support to move, its amplitude at $M_R = 1$ with the transducer
being $0.0211$ against strip theory's $0.0176$, where the pitch differs by two per cent. A
free-free harvester sized with strip theory would therefore be a little under-provisioned in
suspension travel, though not dramatically so.

\section{The compressible section with and without a free support}
\label{sec:versus}

Table~\ref{tab:cfd_comparison} puts the compressible companion paper's rigidly supported
section beside the same section on a fuselage as heavy as the wing, with the aerodynamics held
fixed, so that every difference in it is the fourth degree of freedom and nothing else.

\begin{table}[H]
\centering
\caption{The compressible section on a rigid support, which is the companion paper, against
the same section on a free one.}
\label{tab:cfd_comparison}
\small
\begin{tabular}{lrr}
\toprule
quantity & three degrees of freedom, & four degrees of freedom, \\
 & rigid support & free support, $\MR = 1$ \\
\midrule
baseline flutter boundary $\Us_F$ & 6.288873 & 6.208648 \\
its frequency & 0.083523 & 0.085362 \\
pitch mounting, shift at $\chi = 0.50$, $r = 1$ [\%] & +1.951 & -0.533 \\
limit cycle, pitch amplitude [rad] & 0.0534 & 0.0557 \\
limit cycle, plunge amplitude & 0.1546 & 0.1772 \\
limit cycle, support amplitude & 0.0000 & 0.0211 \\
limit cycle, period & 75.23 & 73.87 \\
mean harvested power $\langle\bar p\rangle$ & 2.431e-05 & 3.283e-05 \\
\bottomrule
\end{tabular}

\end{table}

\textbf{What carries over.} Past findings results are confirmed also here, namely that strip
theory under-predicts the harvested power of a compressible section. It does so by $35.1\%$ on
a rigid support and by $34.9\%$ on a free one, the two agreeing to within two tenths of a
point. The mesh behaves similarly, the transducer at zero coupling leaves the spectrum untouched, and the flow
model's own verification carries over unchanged. The plunge mounting keeps its advantage over
the pitch mounting in both models and on both supports.

\textbf{What the fourth degree of freedom changes.} The boundary falls by $1.276\%$, the
flutter frequency rises, the cycle grows in every coordinate and its period shortens, and the
harvested power rises by $35.1\%$. The support itself takes an amplitude of $0.0211$ where on
a rigid mounting it is structurally zero, which is a quantity the companion paper has no
column for. And the pitch mounting's shift of the boundary at the anchor load resistance goes
from $+1.951\%$ to $-0.533\%$.

\textbf{What only appears with both.} The two effects studied in this series, compressibility
and a free support, cannot be told apart by looking at the rigidly supported compressible
section alone, nor at the incompressible free one. It is only with both present that the shift
at the anchor resistance changes sign, that the band of disagreement between the two
aerodynamic models widens to include that resistance, and that the super-compounding of
Section~\ref{sec:coupling} can be measured at all.

\section{Discussion}
\label{sec:discussion}

\textbf{Whether single-effect studies compose depends on the quantity, and on the regime.}
The two changes studied here were each characterised alone and separately, and each was
reported as roughly halving the pitch mounting's stabilisation at $r = 1$ and as raising the
harvested power by about a third. One of those summaries survives composition and the other
does not. On the harvested power the two are multiplicative to within three tenths of a
percentage point, so a reader holding both companion papers may multiply their corrections and
be right. On the stability of the boundary that same reader, doing the same thing, would
predict a small positive shift where the answer is negative. That failure is not general
either, and Section~\ref{sec:compose} bounds it. Where the load resistance puts the electrical
pole above the flutter frequency the two corrections multiply to within seven parts in a
thousand on both mountings, as cleanly as they do on the power. The failure belongs to the
damping regime alone, which is the regime both companion papers chose to work in, so a reader
is misled in exactly the place the literature invites them to look.

\textbf{Why they differ.} The harvested power is set by the amplitude of the cycle, and both
changes raise that amplitude by making the path between the air and the transducer more
compliant, a compressible flow by loading the section more heavily at a given motion and a
free support by letting the wing reach a given absolute plunge on less compression of its
suspension. Two independent gains of about a third compound as gains do. The transducer's
shift of the boundary at $r = 1$ is not an amplitude at all. It is a phase-dependent damping,
which depends on where the flutter mode sits and on its frequency, and both changes move both,
so there the two are not independent and do not simply multiply. The resistance
sweep is the test of that reading rather than a restatement of it, since the reading predicts
that wherever the transducer stops being a damper the two should compose again, and above
$r = 1.89$ they do.

\textbf{What this costs to compute.} The obstacle a reader would expect, a
four-degree-of-freedom mesh motion, does not exist. A flow solver written for a section on a
rigid support needs no modification whatever to carry a free one, because the wall
displacement conjugate to the support's translation is the wall displacement conjugate to the
plunge. The strip-theory companion paper says as much in passing, and the present work is where
that route is taken, the whole extension being the two maps of Eq.~\eqref{eq:Tmap} at the
provider's interface. That is worth saying plainly, since it makes high-fidelity aerodynamics on a
free-free aeroelastic harvester a matter of hours rather than of writing a new solver.

\textbf{Limitations.} The sign reversal is mapped in load resistance, where it spans at least
a decade and closes where the electrical pole crosses the flutter frequency
(Section~\ref{sec:band}), and in coupling, where the shift changes sign at
about $\chi = 0.315$. The composition of Section~\ref{sec:compose} is mapped at
$\chi = 0.50$ only, and at four resistances rather than a continuum, so the regime boundary it
identifies is located to the decade and not more finely. What is not resolved is the Mach
dependence, the whole study being at $M = 0.10$, nor whether the crossing coupling itself moves
with the load resistance or the support mass. The harvested power is compared at one operating point rather than over the plane, the
full-model marches being the expensive part. No reduced model of the compressible free-free
system was built, so the question the first companion paper asked of a reduced harvester
model is not asked again here. The fuselage remains a point mass in plunge alone, so the model
is a free-free section and not a free-free aircraft, and the flow is inviscid.

\section{Conclusions}
\label{sec:conclusions}

\begin{enumerate}[leftmargin=*,itemsep=2pt]
\item A flow solver coupled to a section on a free support needs no four-degree-of-freedom
mesh motion. The wall displacement conjugate to the support's translation is identically the
one conjugate to the plunge, so the solver still sees three coordinates and the fourth is
carried by a motion map and a force map at its interface. The mesh motion, the residual and
the loads are unchanged.

\item The implementation reproduces the companion compressible paper three times over in the
heavy-fuselage limit, on the flutter boundary, on the pitch-mounted transducer's shift of it,
and on every measured quantity of the full nonlinear limit cycle with the harvester.

\item A compressible flow and a free support each reduce the pitch mounting's stabilising
shift of the flutter boundary where the electrical time constant meets the flutter frequency,
and together they reduce it by \emph{more} than the product of what each does alone. That holds at
both couplings tested. At a coupling of $0.50$ it carries the shift past zero at
the anchor load resistance, where neither change alone does so, and at $0.20$ the shift is
reduced to a tenth of its rigidly supported strip-theory value but stays positive. The super-compounding is the robust finding and the
sign reversal is its consequence at a strong enough coupling.

\item A reversed sign is not itself new, the compressible companion paper having reported one
at a load resistance ten times smaller. What the free support does is widen the band of load
resistance over which the two aerodynamic models disagree in sign, moving its upper edge from
between $r = 0.3$ and $r = 1$ to between $r = 1$ and $r = 3$, so that the band comes to include
the resistance at which the harvesting studies are done. The plunge mounting never changes
sign anywhere in the sweep, on either support and in either model, so the reversal belongs to
the pitch mounting.

\item The failure to compose, however, belongs to both mountings, and it is bounded by the
same resistance that bounds the reversal. Above the resistance at which the electrical pole
meets the flutter frequency, where the transducer acts as an added stiffness, the product of
the two published corrections predicts the measured shift to within seven parts in a thousand
on both mountings. Below it, where the transducer acts as a phase-dependent damper, the
product is wrong by a factor of between one and a half and fourteen, and on the pitch mounting
at the harvesting resistance by a sign. Composability on the boundary is therefore a property
of the regime the transducer is working in, and the regime in which it fails is the one both
companion papers reported their effects in.

\item The same transducer that lowers the boundary flattens the growth-rate curve above it, so
it is destabilising in the sense that matters for where the cycle begins and stabilising in
the sense that matters for how fast it grows. A harvester study should say which it means.

\item On the harvested power the same two effects are multiplicative, giving together
$82.5\%$ against the $82.8\%$ the product of the two alone would predict. At the same load
resistance the boundary does not compose, so the two quantities answer the same pair of
modelling changes differently and a composition rule established for one cannot be carried to
the other.

\end{enumerate}

\section*{Data availability}

All data generated in this study, the scripts that produce every figure and table, and the
source code of the aeroelastic solver are available from the corresponding author upon
reasonable request.

\section*{Funding and competing interests}

The authors received no financial support for this work and declare no competing interests.

\section*{Declaration of generative AI use}

Claude (Anthropic) was used to assist in drafting and editing the text, refining its language
and proofreading it, and in checking the analysis scripts and the consistency of the text with
the computed data. The authors reviewed all content and take full responsibility for it.


\end{document}